\documentclass[useAMS,usenatbib]{mnras}
\usepackage{newtxtext,newtxmath}
\usepackage{xcolor}

\newcommand{\Msun}{M$_{\odot}$}
\newcommand{\kms}{km\,s$^{-1}$}

\title[The wide MSTO and upper MS of NGC 1831]{The wide upper main sequence and main sequence turnoff of the $\sim$\,800 Myr old star cluster NGC 1831\thanks{Based on observations with the
NASA/ESA {\it Hubble Space Telescope}, obtained at the Space Telescope Science
Institute, which is operated by the Association of Universities for Research in
Astronomy, Inc., under NASA contract NAS5-26555}}

\author[M. Correnti et al.]{Matteo Correnti$^{1}$\thanks{E-mail: correnti@stsci.edu}, Paul Goudfrooij$^{1}$, Andrea Bellini$^{1}$, and Leo Girardi$^{2}$\smallskip \\ 
$^{1}$Space Telescope Science Institute, 3700 San Martin Drive, Baltimore, MD 21218, USA\\
$^{2}$INAF\,--\,Osservatorio Astronomico di Padova, Vicolo dell'Osservatorio 5, I-35122, Padova, Italy\\}

\usepackage{graphicx}

\pagerange{\pageref{firstpage}--\pageref{lastpage}} \pubyear{xxxx}

\begin{document}

\date{Accepted 2021 January 21. Received 2020 December 22; in original form 2020 July 23}

\maketitle

\label{firstpage}

\begin{abstract}
We present the analysis of the colour-magnitude diagram (CMD) morphology of the $\sim$ 800 Myr old star cluster NGC\,1831 in the Large Magellanic Cloud, exploiting deep, high-resolution photometry obtained using the Wide Field Camera 3 onboard the {\it Hubble Space Telescope}. We perform a simultaneous analysis of the wide upper main sequence and main sequence turn-off observed in the cluster, to verify whether these features are due to an extended star formation or a range of stellar rotation rates, or a combination of these two effects. Comparing the observed CMD with Monte Carlo simulations of synthetic stellar populations, we derive that the morphology of NGC\,1831 can be fully explained in the context of the rotation velocity scenario, under the assumption of a bimodal distribution for the rotating stars, with $\sim$ 40\% of stars being slow-rotators ($\Omega / \Omega_{\rm crit} < 0.5$) and the remaining $\sim$ 60\% being fast rotators ($\Omega / \Omega_{\rm crit} > 0.9$). We derive the dynamical properties of the cluster, calculating the present cluster mass and escape velocity, and predicting their past evolution starting at an age of 10 Myr. We find that NGC\,1831 has an escape velocity $v_{esc} = 18.4$ \kms\, at an age of 10 Myr, above the previously suggested threshold of 15 \kms\, below which the cluster cannot retain the material needed to create second-generation stars. These results, combined with those obtained from the CMD morphology analysis, indicate that for the clusters whose morphology cannot be easily explained only in the context of the rotation velocity scenario, the threshold limit should be at least $\sim$ 20 \kms. 
\end{abstract}


\section{Introduction}
\label{s:intro}

In the last dozen years, high-quality colour-magnitude diagrams (CMDs) taken with the Advanced Camera for Survey (ACS) and the Wide Field Camera 3 (WFC3) on board the {\it Hubble Space Telescope (HST)} revealed that the majority of massive intermediate-age star clusters (1\,--\, 2 Gyr old) in the Magellanic Clouds (MC) exhibit extend main sequence turn-off (hereafter eMSTOs) that extend both in brightness and colour \citep{mack+08a,milo+09,goud+09,goud+11a}. In addition to eMSTOs, several of these clusters have been found to exhibit striking dual clumps of red giants, where the main clump of core helium burning stars is followed by a faint extension (or secondary clump) containing $\sim$ 10\,--\,20\% of the clump stars \citep{gira+09,gira+13, rube+11}. More recently, eMSTOs have been observed also in some young (100\,--\,300 Myr old) star clusters of the LMC, where they are often associated with the presence of a split MS \citep[e.g.,][]{milo+15,milo+17,corr+17}. 

Since the discovery of eMSTOs in star clusters, two main opposing scenarios have been put forward to explain such features, and the nature of this phenomenon is still debated. One interpretation is that eMSTOs are due to stars that formed at different times within the parent cluster, with spread in age of 100-500 Myr \citep[i.e., age spread scenario;][]{mack+08a,milo+09,corr+14,goud+14,goud+15}. \citet{goud+14} postulated that in this scenario, the shape of the star density across the eMSTO reflects the combination of two effects: the histories of star formation of those clusters and the mass loss due to the cluster expansion after the death of the massive stars in the innermost regions. In this context, \citet{goud+14} found a correlation between the width of the MSTO and the cluster central escape velocity, which can be considered a key element for the cluster ability to retain and/or accrete gas and stellar ejecta during the early evolution phase. These findings, combined with the strong correlation discovered by \citet{goud+14} between the fractional number of stars in the bluest region of the MSTO and those in the faint extension of the red clump, provided observational support for the age spread scenario.

The other main interpretation, originally proposed by \citet{basdem09}, is the ``stellar rotation scenario'' where eMSTOs are due to the presence of a spread in rotation velocities among turn-off stars of a coeval population \citep[for a detailed discussion, see][and references therein]{gira+11,goud+14,geor+14,nied+15,brahua15,dant+18,geor+19}. The aforementioned discovery of  MS splits in young LMC clusters provided significant support for the stellar rotation scenario, since the Geneva SYCLIST model for stellar rotation in isochrones \citep{geor+14} predicts broadened or split MSs at these ages when a large fraction of stars has high rotation rates (in excess of $\sim$ 80\% of the critical rate $\Omega_{C}$), whereas wide MSs cannot be easily explained by age spreads \citep{dant+15,milo+16,corr+17}.   
Moreover, it was observed that several stars in the MSTOs of these young massive clusters are strong $H\alpha$ emitters and are thought to constitute a population of rapidly rotating ($\Omega / \Omega_{C} \geq$ 0.5) equator-on Be stars \citep{bast+17,corr+17,dupr+18,milo+18}. This scenario has been further strengthened by the evidence found through spectroscopic studies: \citet{mari+18b} found that in the young cluster NGC\,1818 the MS splits in two components, with the blue MS hosting slow rotators and the red MS hosting fast rotators, respectively. \citet{kama+20} studied the intermediate-age ($\sim$ 1.5 Gyr old) star cluster NGC\,1846 and discovered a bimodal distribution in the rotation rates of the MSTO stars, with a fast component, centered around $\nu$ sin{\it i} = 140 \kms, and a slow component centered around $\nu$ sin{\it i} = 60 \kms.\\ Recent observations of Milky Way open clusters with \textit{GAIA} \citep{mari+18a,cord+18,bast+18} have revealed that eMSTOs are common also in those objects, indicating that this feature may be a natural property of young clusters, rather than a peculiarity of MC clusters.\\
Finally, another piece of evidence that seems to favour the stellar rotation scenario is that this scenario predicts that, if the extension of the MSTO due to rotation is interpreted as an age spread, then the inferred age spread should increase as a function of the age of the cluster \citep[][]{nied+15}. This seems to be confirmed by observational and theoretical studies that show how the extension of the MSTO appears to correlate with the cluster age, at least up to an age of $\sim$\,1 Gyr   \citep[][and references therein]{bast+18,geor+19}.

All these results point towards the conclusion that stellar rotation plays a fundamental role in shaping the morphology of eMSTOs.
However, despite this evidence, there are indications that other effects can be at play as well. For example, the distributions of stars in the eMSTO of several massive young clusters are not consistent with a coeval population of stars with different rotation rates, as represented by the SYCLIST model predictions \citep[see for example][]{milo+16,milo+17,milo+18,corr+17}. Moreover, \citet{goud+17} demonstrated that the distribution of stars across the eMSTO of two massive clusters with age $\sim$ 1 Gyr cannot be explained solely by a distribution of rotation rates according to the SYCLIST models, unless the orientations of rapidly rotating stars are heavily biased toward an equator-on configuration in both clusters. Finally, the mechanisms that lead to the presence and the evolution of the rotation rates in these clusters are still uncertain \citep[see, e.g.,][and references therein]{dant+17}.
 
One of the most prominent complications in these studies is that the SYCLIST stellar tracks that incorporate the effects of rotation are only available for stars with M $>$ 1.7 \Msun, while the MSTO stars in the massive intermediate-age clusters studied so far are less massive due to their ages. To circumvent these problems, it is fundamental to study star clusters that have the following properties: a) they need to be young enough for the MSTO and upper MS regions to consist of stars with M $>$ 1.7 \Msun\, thus with an age $<$ 800 Myr; b) they need to be massive enough to enable statistically significant measurements of the morphologies and extents of the MSTO and upper MS regions in the CMD (i.e., the cluster should have a mass of several $10^4$ \Msun); c) more importantly, they need to have a (mean) age at which the CMD morphologies for the MSTO and upper MS are predicted to be substantially different for the two scenarios. This is the case of clusters with ages $>$ 500 Myr. In particular, the magnitude extent of the MSTO in colour is similar to that in brightness for the case of an age range, while it is larger in brightness than in colour for the case of a range in rotation velocities.  Conversely, the upper MS is expected to be consistent with a single-age simple stellar population for the age range case, whereas a range in rotation velocities causes a significantly wider MS.

With this in mind, we analyze new deep WFC3 and ACS {\it HST} observations of the cluster NGC\,1831. NGC\,1831 satisfies all the requirements listed above; it has an age ($\sim$ 800 Myr, and hence a MSTO mass $M_{\it MSTO} \simeq 1.9$ \Msun), and mass 
\citep[$\sim 4\times 10^4 \: M_{\odot}$, see][]{mclvan05} for which the morphologies of the upper MS and MSTO in CMDs of selected passbands allow a direct discrimination between ranges of ages versus ranges of stellar rotation rates. The CMD of the inner region of the cluster has already been presented in \citet[][hereafter G18]{goud+18}, which reported the presence of an eMSTO and focused their analysis on the kink feature observed in the cluster MS and its implications on the nature of the eMSTOs observed in young and intermediate age star clusters. Moreover, \citet[][hereafter G19]{goss+19} analyzed the MSTO of NGC\,1831, comparing the observed CMD with synthetic stellar populations based on MESA stellar models. In their analysis, they found comparable statistical evidence between rotation and an age spread
in fitting the observed eMSTO structure. 

Here, we present a complete study of the CMD morphology, performing a detailed and simultaneous analysis of the eMSTO and upper MS regions, comparing the cluster CMD with Monte Carlo simulations of synthetic star cluster with multiple single stellar populations (SSP) of different ages and with a single age SSP with a range of different stellar rotation velocities. This study allows us to provide a better picture 
with regard to the causes related to the nature of the eMSTO phenomenon.

The remainder of the paper is organized as follows: we present observations and data reduction in Section\,2. In Section\,3, we discuss the observed CMD, whereas in Section\,4 we compare the observed CMD with theoretical models, exploring the two different scenarios through the use of isochrones with different ages and different rotation rates. Section\,5 shows the comparison with Monte Carlo simulations, both using SSP with different ages and different rotation rates; from the comparison between observed and simulated pseudo-colour distributions we find the best-fit models. The physical and dynamical properties are presented in Section\,6. Finally, we discuss our results and their implications on the interpretation of the eMSTO nature in Section\,7.

\section{Observations and Data Reduction}
\label{s:data}
Observations of the cluster NGC\,1831 were obtained on 2016 November 21 using the UVIS channel of WFC3 as part of the {\it HST} program 14688 (PI: P. Goudfrooij). As in previous programs, we centered the cluster on one of the two CCD chips of the WFC3/UVIS camera, so that the observations cover enough radial extent to allow a useful King-model fit to the radial stellar density profile. Moreover, with this choice the full effective radius of the cluster fits within one chip, minimizing the impact of chip-dependent zeropoints, and avoiding the loss of the cluster central region due to CCD chip gap. Observations were performed in two filters, {\em F336W} and {\em F814W}. The choice of the filter {\em F336W} as the blue filter provides the best compromise between a reasonable exposure time and the ability to observe well-detectable differences in colour between stars with different rotation rates. Four long exposures were taken in the {\em F336W} filter (total exposure time 4180 sec), and two in the {\em F814W} filter (total exposure time 1380 sec). In addition, we took a short exposure of 100 sec for the filter {\em F814W} to avoid saturation of the brightest stars. A spatial offset between the different exposures was applied in order to cover the gap between the two WFC3/UVIS chips and to facilitate the identification and removal of hot pixels. We also used the Wide Field Channel (WFC) of the Advanced Camera for Surveys (ACS) in parallel mode to obtain images $\sim$ 6\arcmin\ away from the cluster center, using filters {\em F435W} and {\em F814W}. These observations provide a clean picture of the stellar content of the underlying LMC field, allowing us to establish the true background level and resulted to be fundamental for the determination of the cluster structural parameters and dynamical properties. Observation details are summarized in Table\ref{t:obs}

\begin{table}
\begin{center}
\begin{tabular}{ccc}
\hline
\hline
Camera & Exposure Time & Filter\\ 
 \hline
WFC3/UVIS & 2$\times$975 s $+$ 2$\times$1115 s & {\em F336W}\\
 & 100 s $+$ 660 s $+$ 720 s &  {\em F814W}\\
 ACS/WFC & 850 s $+$ 890 s $+$ 980 s & {\em F435W}\\
  & 600 s $+$ 950 s & {\em F814W}\\
\hline 
\end{tabular}
\caption{Summary of NGC\,1831 observations}
\label{t:obs}
\end{center}
\end{table}

To reduce the images, we follow the same procedure outlined in \citet[][hereafter C17]{corr+17} and described in more detail in \citet{bell+17}. First, calibrated \texttt{.flc} science images are retrieved from the {\it HST} archive. The \texttt{.flc} images are the products of the \texttt{calwf3} data reduction pipeline and constitute the bias-corrected, dark-subtracted, flat-fielded, and charge transfer inefficiency corrected images. Then, we preliminary measure star positions and fluxes using the publicly available \texttt{FORTRAN} program {\em img2xym\_wfc3uv} with the available library of spatially variable point-spread functions (PSFs, in an array of $7 \times 8$ PSFs) for each filter. To account for telescope breathing effects \citep{bell+14}, an additional $5 \times 5$ array of perturbation PSFs for each exposure is derived and combined with the library PSFs to perform the final stellar profile fit. We correct the stellar positions for geometric distortion using the solution provided by \citet{belbed09,bell+11}. Using one of the long {\em F814W} exposures, we create a master-frame list and we transform and average the other filter exposures into the reference system defined by the master frame. To do so, we use a six-parameter linear transformation. Our final catalog of stellar positions and fluxes is obtained through simultaneous fitting of all the available exposures using the FORTRAN software package KS2 \citep[see][for a detailed description of the code]{bell+17}. KS2 allows us to measure faint objects that would have been otherwise lost in the background noise of single exposure. In addition, KS2 performs several waves of source finding and subtraction to carefully deblend the contribution of neighbors. Photometry for saturated stars is derived using the method developed for the ACS camera by \citet{gill04} and subsequently demonstrated to be valid also for the WFC3/UVIS detector \citep{gill+10}. This method allows us to recover the electrons that have bled into neighbouring pixels.

Photometric calibration is performed by comparing the PSF-based instrumental magnitudes, measured on the \texttt{.flc} exposures, with aperture-photometry magnitudes measured on the \texttt{.drc} exposures, using a fixed aperture radius of 10 pixels. The \texttt{.drc} exposures, constitute the drizzled, stacked, distortion-corrected, and flux-normalized images, obtained by combining together the individual \texttt{.flc} images using the \texttt{Drizzlepac} software \citep{gonz+12}. Briefly, we first align all the images in the same filter using the software \texttt{TweakReg} and then, using the software \texttt{AstroDrizzle}, bad pixels and cosmic rays are flagged and rejected from the input images. Finally, undistorted and aligned images are combined together into a final stacked image. Magnitudes are transformed into the \texttt{VEGAMAG} system adopting the relevant synthetic zero-points\footnote[3]{https://www.stsci.edu/hst/instrumentation/wfc3/data-analysis/photometric-calibration/uvis-photometric-calibration}. In the final catalog we retain only the sources that are recovered in both exposures for the two filters. Furthermore, the catalog is cleaned from spurious detections using the different diagnostics provided by \texttt{KS2}, which allow us to determine the quality of the PSF model fit for each source \citep{bell+17}.  

\begin{figure}
\includegraphics[width=1\columnwidth]{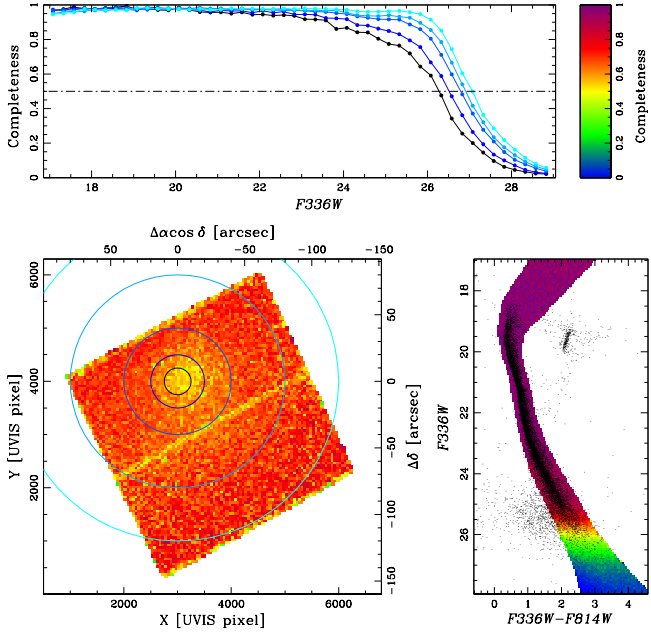}
\caption{Bottom panels: completeness fraction as a function of only spatial position and only colour-magnitude relation (with superimposed the cluster CMD), colour-coded as shown in the top-right panel (i.e., from red to blue for a completeness fraction from 1 to 0). Top-left panel: average completeness fraction of the artificial star test experiments in the central region of the cluster and four different annuli. The five different region are shown in the bottom-left panel, superposed on the XY spatial distribution.}
\label{f:completeness}
\end{figure}

To characterize the completeness and the photometric-error distribution of the final photometry, we perform artificial star tests, using the standard technique of adding artificial stars to the images and then reducing them using the same procedure as for the real data \citep[for a detailed description, see][]{bell+17}. Here, we generate a list of $1 \times 10^{6}$ artificial stars and we inject them following the overall distribution of the real stars in the image. Artificial stars are added one at a time so that they do not interfere with each other and do not artificially alter the crowding of the image and are distributed in magnitude and colour intervals that cover the observed CMD range. The inserted artificial star is considered recovered if the input and output magnitude agree within 0.75 mag in both filters and if the input and output position agree to within 0.5 pixel. Finally, we assign a completeness fraction to each individual star in the CMD as a function of its magnitude and distance from the cluster center. The completeness fraction as a function of spatial position (regardless of input magnitude) and as a function of magnitude and color (regardless of input position) are reported in the bottom-left and bottom-right panels of Fig.\,\ref{f:completeness}, following the colour coding reported on the top-right panel. Average completeness fractions as a function of the {\em F336W} magnitude for a region centered on the cluster and four different annulii (superposed on top of the spatial distribution reported in the bottom-left panel) are shown in the top-left panel with different colours.

\section{Colour-Magnitude Diagram Analysis}
\label{s:cmd}

\begin{figure}
\includegraphics[width=1\columnwidth]{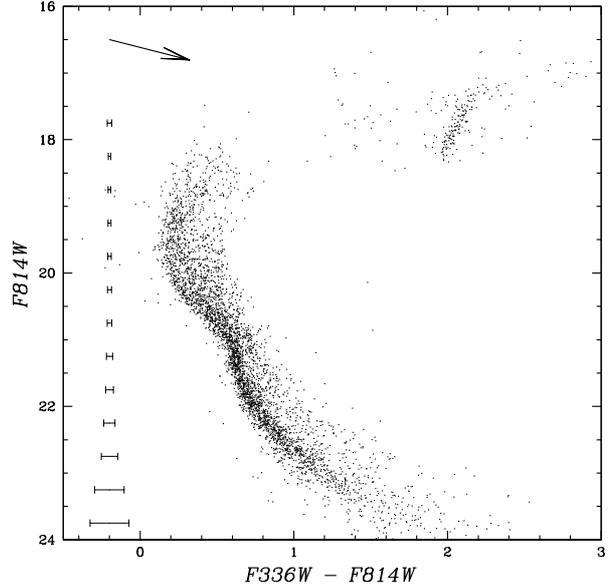}
\caption{{\em F814W} versus {\em F336W - F814W} CMD for all the stars inside the NGC\,1831 field, defined in Section\,\ref{s:king}. Magnitude and colour errors are shown in the left-hand side of the CMD. We also report the reddening vector for $A_V$ = 0.5.}
\label{f:cmd}
\end{figure}

Fig.\,\ref{f:cmd} shows the {\em F814W} versus {\em F336W - F814W} CMD of the cluster field, defined as the region enclosed within the projected effective radius of the cluster ($r_e$ = 35.2 arcsec, see Section\,\ref{s:king}). As shown in Fig.\,1 of G18, the contamination by field stars is negligible and does not affect the features observed in the cluster CMD.  The cluster CMD can be split in two portions, each characterized by a different morphology. The two portions are above and below the kink at {\em F814W} $\sim$ 21 mag (see Section\,\ref{s:iso} and G18 for a detailed description of this feature). Above the kink, the MS and MSTO are wider than what is expected from a SSP, whereas below the kink, the shape of the single star MS resembles the SSP one. In fact, from the comparison with a synthetic cluster of a simple stellar population, G18 noted that the cluster MSTO is significantly more extended than the SSP one in conjunction with photometric uncertainties.

As a final check, we verify whether differential reddening is present in our field of view, with a particular focus on the possible influence to the cluster morphology in the upper MS region. We adopt the same approach described in detail in our previous papers \citep[][hereafter C15, C17, and references therein]{corr+15}.

\begin{figure*}
\centerline{
\includegraphics[height=8.2cm]{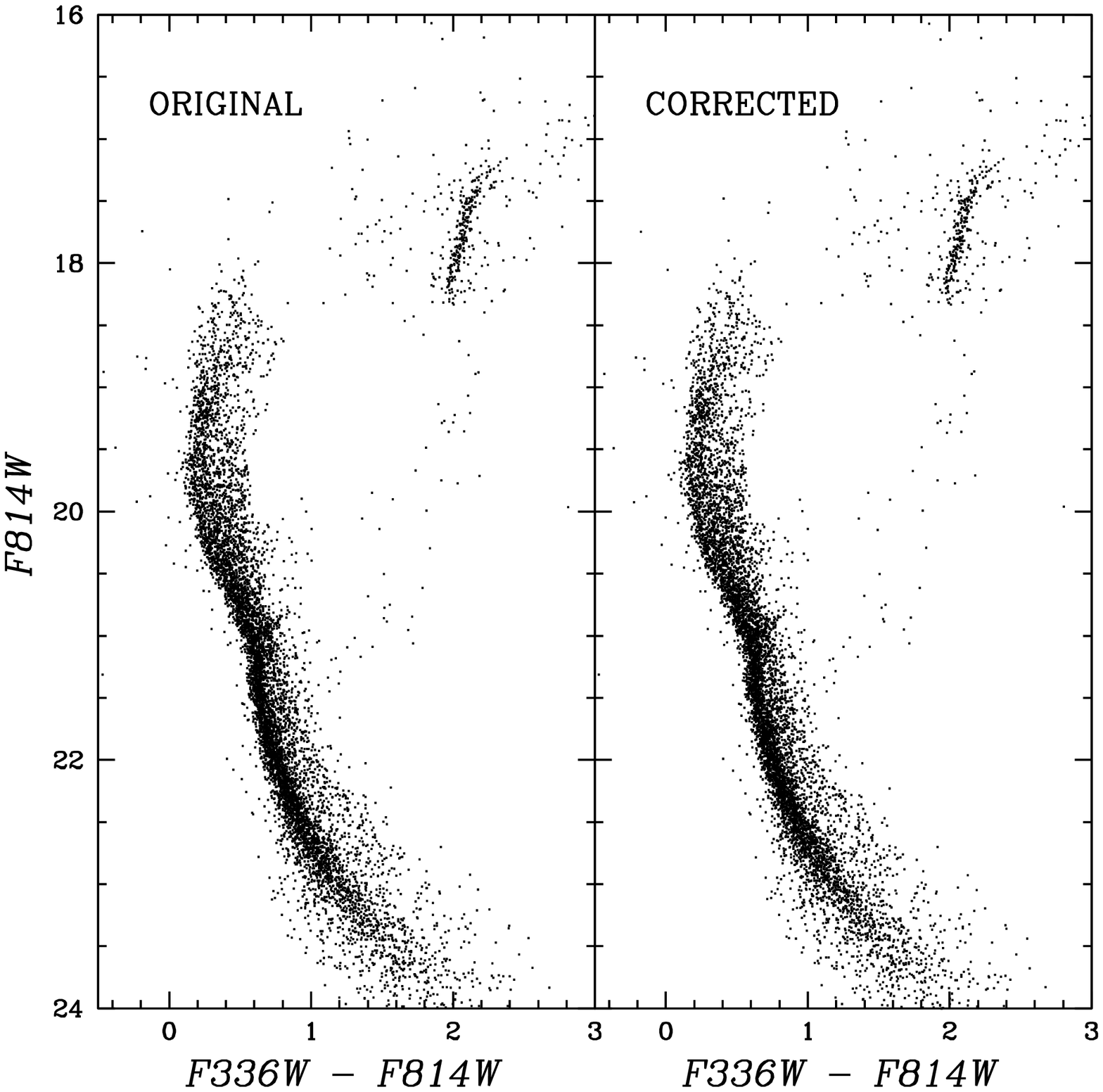}
\includegraphics[height=8.2cm]{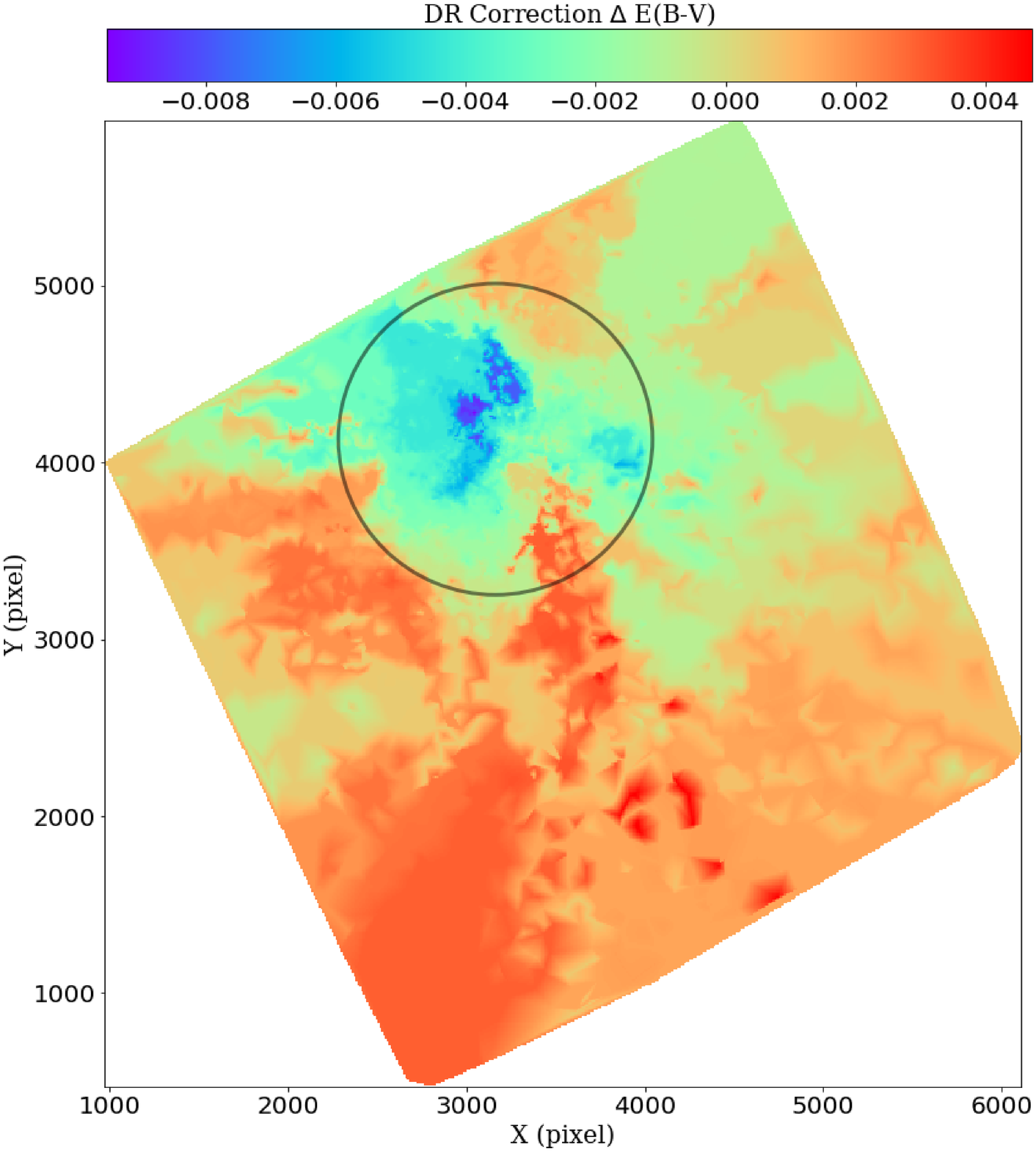}
}
\caption{Left hand-panels: original {\em F814W} versus {\em F336W - F814W} CMD in the WFC3/UVIS field of view compared with the CMD corrected for differential reddening. Right-hand panel: spatial distribution of the differential reddening, colour-coded depending on the differential reddening correction, as shown at the top. The black circle marks the effective radius of NGC\,1831 (i.e., r = 35.2 arcsec).}  
\label{f:redmap}
\end{figure*}

The results of the differential reddening correction are shown in Fig.\,\ref{f:redmap}, where we compare the original CMD (left-hand panel) and the CMD corrected for differential reddening (middle panel). The differences between the two CMDs are negligible as expected given the overall very low extinction in the direction of the cluster and the relatively small field of view of our data. As a further demonstration, in the right panel of Fig.\,\ref{f:redmap} we show the spatial distribution of the differential reddening in our field of view. We report each star with a different colour, depending on the final {\it E(B-V)} that we apply, and we identify with a circle the region defined as the cluster field (i.e., $r < r_e$). The spatial distribution clearly shows that reddening variations are very small, of the order of $\pm 0.01$ mag, thus confirming that differential reddening do not affect the CMD morphology. Hence, in the following analysis we use the original photometry for simplicity. 

\section{Isochrone fitting}
\label{s:iso}

\begin{figure*}
\begin{center}
\includegraphics[height=12cm]{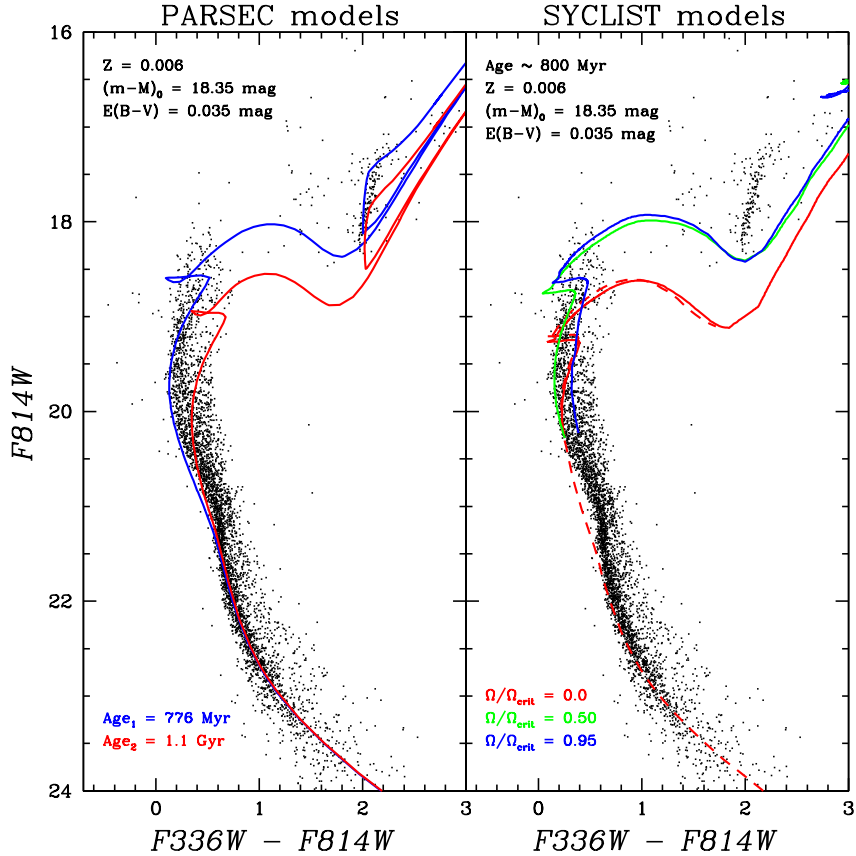}
\caption{Left panel: {\em F814W} versus {\em F336W - F814W} CMD with superimposed best-fitting isochrones from PARSEC \citep{bres+12} for the minimum (blue line) and maximum (red line) age (776 Myr and 1.1 Gyr, respectively) that can be accounted for by the data. The derived best-fit metallicity, distance, and reddening are reported in the figure. Right panel: Same CMD as in the left panel, but with superposed isochrones with different rotation rates from the SYCLIST Geneva database \citep{geor+14}. Isochrones are colour-coded depending on their $\Omega / \Omega_{\rm crit}$ as follows: non-rotating isochrone ($\Omega / \Omega_{\rm crit} = 0.0$), red line, intermediate-rotating stars isochrone ($\Omega / \Omega_{\rm crit} = 0.50$), green line, and fast-rotating stars isochrone ($\Omega / \Omega_{\rm crit} = 0.95$), blue line. The non-rotating isochrone from \citet{mowl+12}, used to extend the $\Omega / \Omega_{\rm crit} = 0.0$ isochrone below M $<$ 1.7 \Msun\, is depicted as a dashed red line. The adopted age, metallicity, distance and reddening are also reported.} 
\label{f:cmd_iso}
\end{center}
\end{figure*}

As a first step to understand the nature of the features observed in the cluster MSTO and upper MS regions, we compare the CMD of NGC\,1831 with model isochrones, using the same approach adopted in C17. Thus, we compare the cluster CMD with a set of non-rotating isochrones, having different ages, and with a set of single-age isochrones but with different rotation rates. The two different comparisons are shown in Figure\,\ref{f:cmd_iso}.

The left panel of Fig.\,\ref{f:cmd_iso} shows the cluster CMD with two superposed isochrones from PARSEC \citep{bres+12} of different ages (776 Myr and 1.1 Gyr). The ages of the isochrones are chosen in order to match the minimum and maximum age that can be accounted for by the position and width of the MSTO. To derive the best-fit isochrone parameters, we use the method described in detail in \citet{goud+09} and adopted in all our previous works. In particular, we adopt the isochrones whose parameters provide a best match to the brightness and colour of core helium-burning stars and the MSTO. To do so, we first retrieve a set of isochrones with a range of metallicity consistent with literature values for the LMC clusters (i.e., Z $\sim$ 0.005\,--\,0.008), separated by step $\Delta Z =$  0.001, and with an age range that encompasses the quoted cluster age of $\sim$ 800 Myr. Then, we select the isochrones for which the values of the difference in magnitude and colour between the MSTO and the red clump (RC) lie within 2$\sigma$ of those parameter observed from the CMD. For the isochrones that satisfy this selection criteria, we find the best-fitting values for distance modulus and reddening by means of a least-squares fitting program to the magnitudes and colours of the MSTO and the RC. Finally, we choose the best-fitting one by means of a visual examination after overplotting the isochrones on the CMD. 

In the right panel of Fig.\,\ref{f:cmd_iso}, we superpose on the cluster CMD the isochrones from the Geneva SYCLIST database \citep{ekst+12,geor+13,geor+14} for which different rotation rates are available. To keep the figure clear and readable, we report only three isochrones with different rotation rates: the red isochrone represents no rotation, i.e. $\Omega / \Omega_{\rm crit} = 0.0$, where $\Omega_{\rm crit}$ denotes the critical (breakup) velocity, the green isochrone represents intermediate-velocity rotators ($\Omega / \Omega_{\rm crit} = 0.5$), whereas the blue isochrone identify fast-rotators stars ($\Omega / \Omega_{\rm crit} = 0.95$). The non-rotating isochrone is extended to masses M $<$ 1.7 \Msun\, by using the model of \citet[][reported in the CMD as a red dashed line]{mowl+12}. To account for the small colour mismatch present between the two non-rotating isochrones, due to slightly different input physics in the models, we apply a small colour shift (i.e., $\sim$ 0.06 mag) to the \citet{mowl+12} model. We note that the isochrones from the SYCLIST database are not able to reproduce the observed morphology of the RC of core helium-burning stars, hence in the following we will limit our analysis to the upper MS and MSTO regions. Moreover, since we could not use the RC as a lever arm in the  determination of population parameters for these set of isochrones, we adopt the same parameters, in terms of metallicity\footnote{Note that the Geneva database for rotating isochrones provides only three options for metallicity, and we choose the most appropriate for our case (i.e., $Z$ = 0.006).}, distance, and reddening, derived from the PARSEC isochrone-fitting and we choose the age that provides the best-fit to the observed CMD. 

Both non-rotating isochrone models fail to provide a satisfactory fit in the region of the MS kink at {\em F814W} $\sim$\,21. The causes related to the discrepancy between observations and models in this region of the MS are described in detail in G18. Briefly, G18 suggested that the position and location of the MS kink observed in young LMC star clusters is associated with a sudden change in the extent of the convective envelope, at the metallicity of these objects, with a sudden onset of strong convection in the outer layers of the stars, and that represents an empirical measure of the stellar mass below which rotation has no appreciable influence on the energy output of the star. In fact, the location of the kink corresponds to an initial stellar mass  M/\Msun\ $\simeq$ 1.45 that, in turn, corresponds to the mass range where stellar structure are subject to significant changes between radiative and convective modes. In the Geneva models, stars with M/\Msun\  $\leq$ 1.7 are treated as having non-negligible convective envelopes  featuring magnetized winds that may shed the angular momentum built up by the star during its formation period \citep{geor+14}. Hence, the observed MS kink can be considered an empirical identification of the transition between non-rotating and rotating stars. This seems to be confirmed by the good fit provided by the non-rotating isochrones  for the MS region below the kink (i.e., the region where stars are thought to be non-rotating). 


Finally, the comparison shown in Fig.\,\ref{f:cmd_iso} suggests that both a spread in age and a range of rotation velocities provide an overall good fit for the data. A more detailed analysis, involving the comparison of the observed CMD with Monte Carlo simulations of synthetic star clusters, is presented below. 

\section{Monte Carlo Simulations}
\label{s:simulation}

To determine the causes of the observed cluster morphology, we conduct different Monte Carlo simulations of synthetic star clusters with multiple SSP of different ages and with a single age SSP with a range of different stellar rotation velocities. In the first case, we simulate a spread in age, populating different SSP using a set of PARSEC isochrones, with the minimum and maximum ages chosen to match the values reported in Fig.\,\ref{f:cmd_iso}. To do so, stars for each SSP are randomly drawn using a Salpeter mass function and the total simulated number is normalized to the observed (completeness-corrected) number of stars. A component of unresolved binary stars, derived from the same mass function, is added to a fraction of the sample stars. We use a flat distribution of primary-to-secondary mass ratios and adopt a binary fraction of $\sim$ 20\%. As in the previous works, we verify that the choice of binary fraction, within $\pm 20\%$ does not affect the results. Finally, we add photometric errors with a distribution derived during the photometric reduction process. In particular, for each star, we randomly extract the magnitude errors from a Gaussian with a $\sigma$ defined as the uncertainty at that particular magnitude derived from the artificial star tests. 

To simulate a SSP with a range of rotation velocities, we retrieve cluster simulations from the SYCLIST database. We use the same binary fraction and photometric errors as described above. The simulations also include gravity and limb darkening effects \citep{esplar11,clar00} as well as a random distribution of inclination angles. We adopt the rotation velocity distribution as derived in \citet{huang+10}. 

The comparison between the observed MSTO and upper MS regions and the simulated ones, is carried out using the same approach as in C17 \citep[see][for a detailed description]{goud+11a}. Briefly, we create a series of pseudo-colour distributions, constructing a parallelogram across the MSTO and the upper MS and selecting the stars within. The parallelogram axes are approximately parallel and perpendicular to the isochrones, respectively. For this reason, the distributions are defined with the term 'pseudo-colour' because they reflect the star distribution along the major axis of the parallelogram, rather than the X - Y colour. The position and width of the MSTO parallelogram are chosen such that they include the MSTO stars for all the different populations simulated (i.e., MSTO stars with different ages and MSTO stars with different rotation rates), based on the MSTO locations of the isochrones depicted in Figure\,\ref{f:cmd_iso}, in order to avoid any selection bias. For what concerns the MS parallelogram, we choose an arbitrary width, large enough to provide a good statistical sample (note that changing the width by $\pm$ 0.05 mag does not change significantly the results presented below) and a position that properly takes into account the binary stars in the crosscut. For what concerns the SYCLIST simulations, we use a crosscut perpendicular to the mean orientation of the stars in the cluster
simulation (i.e., perpendicular to the rotating-isochrones shown in Fig.\ref{f:cmd_iso}). Finally, to calculate the pseudo-colour distributions we use the non-parametric Epanechnikov-kernel density function \citep{silv86}, in order to avoid possible biases that can arise if fixed bin widths are used. This procedure is applied to both the observed CMD and the different simulations. In the following sections, we describe the results obtained from the comparison of the observed and simulated pseudo-colour distributions. 

\subsection{Testing the age spread scenario}
\label{s:sim_age}

\begin{figure}
\includegraphics[width=1\columnwidth]{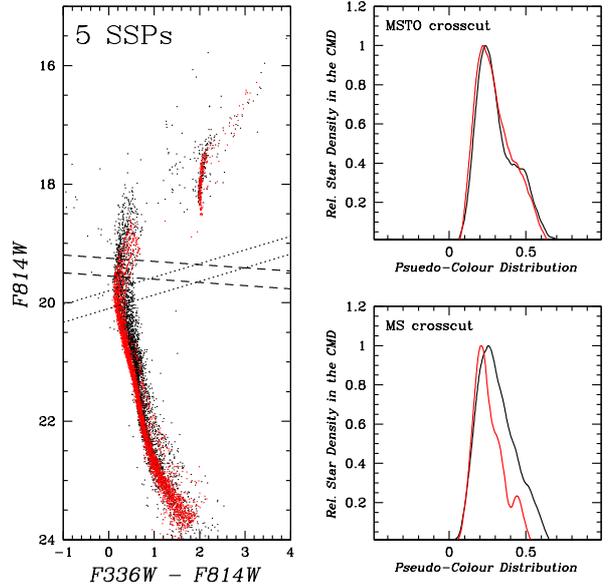}
\caption{Left panel: comparison between the observed (black dots) and simulated (red dots) CMDs. The latter is obtained by combining Monte Carlo simulations of five SSP with different ages, derived populating the PARSEC isochrones as described in Section\,\ref{s:simulation}. We also report the two crosscuts used to select the MSTO and upper MS stars. Right panels: pseudo-colour distributions for the MSTO region (top panel) and upper MS region (bottom panel) of the observed (black line) and simulated (red line) CMDs.}
\label{f:pseudo_age}
\end{figure}

To test whether the observed cluster morphology can be explained in the context of the age spread scenario, we build Monte Carlo simulations combining a set of SSPs, obtained as described above, with different ages. We use five different SSPs, with ages between the minimum and maximum ages derived from the isochrone fitting, using an age step of $\sim$ 70 Myr. To derive the final simulation, we associate a probability {\it p} with each SSP and we extract stars from each SSP according to their associated probability {\it p}. As a first guess, we started with a uniform probability (i.e., {\it p} = 0.2); after inspection of the results, we modified the probability of the oldest SSP to better reproduce the observed CMD morphology, in particular in the MSTO region (i.e., {\it p} = 0.1 for the oldest SSP and {\it p} = 0.225 for the other four). The comparison between the observed CMD and the Monte Carlo simulation is shown in the left panel of Figure\,\ref{f:pseudo_age}. The crosscuts adopted to select the MSTO and upper MS stars are also reported. As expected from the isochrone comparison described in the previous section, this simulation seems to reproduce the full extent of the MSTO, in particular its colour distribution, whereas it fails to fully capture the width of the observed MS region. This is more evident from the comparison of the pseudo-colour distributions, shown in the right panels of Figure\,\ref{f:pseudo_age} (top panel: MSTO crosscut; bottom panel: upper MS crosscut). Indeed, while for the MSTO region the simulated distribution (red line) provides a very good fit of the data, in the upper MS region the observed distribution is wider than the simulated one, extending towards significantly redder colours. 

It is also worth pointing out that in this comparison, we are completely excluding the presence of rotating stars in our data, despite the evidence of the role played by rotation in shaping the cluster morphology of young LMC clusters and MW open clusters. Hence, including the effects of rotations for the different SSP can alter the results obtained here, in particular in the pseudo-colour distribution analysis. However, the purpose of this test is just to establish whether a spread in age alone can reproduce the features observed in the CMD. In this context, it is worth to note for example, that the simulated CMD does not provide a good fit for the blue bright portion of the MSTO (around {\em F814W} $\sim$ 18.5, {\em F336W - F814W} $\sim$ 0.2), where we expect to find the population of fast-rotating stars (see Figure\,\ref{f:cmd_iso} and Section\,\ref{s:sim_rot} below), thus providing further evidence that an age spread alone cannot explain the observed CMD morphology.


\subsection{Testing the stellar rotation scenario}
\label{s:sim_rot}

\begin{figure}
\includegraphics[width=1\columnwidth]{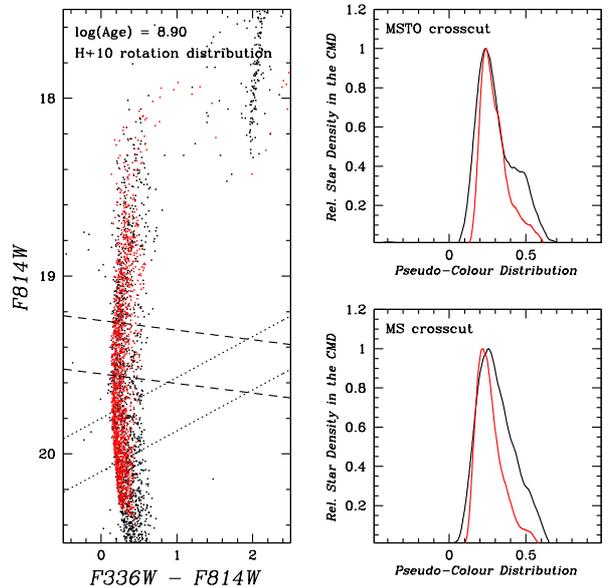}
\caption{Zoom-in of the upper MS region with overplotted on the cluster CMD the Monte Carlo simulation retrieved through the SYCLIST website, having a single age and different rotation velocities. The rotation velocities distribution is from \citet{huang+10}.}
\label{f:pseudo_rot_h10}
\end{figure}

To test the stellar rotation scenario, we compare the Monte Carlo simulation retrieved through the SYCLIST website with the observed CMD. Similarly to Figure\,\ref{f:pseudo_age}, in Figure\,\ref{f:pseudo_rot_h10} we show the cluster CMD and overplot the SYCLIST simulation and the derived pseudo-colour distributions. The analysis of the two CMD morphologies shows that the brightness and colour extension of the MSTO seem to be well reproduced by the simulation, while the observed width of the upper MS seems to be slightly larger than the simulated one. However, it is equally important to take into account not only the actual extension of the feature, but also how the stars are actually distributed in colour inside the feature  (i.e. the ratio of stars in the different regions of the selected cross-cut). In fact, the pseudo-colour distributions shown in the right panels of Figure\,\ref{f:pseudo_rot_h10} indicate that the red side of the simulation is not as highly populated as the observed one and thus it does not reproduce the observed distribution in the MSTO and MS regions. 

It is important to keep in mind that this discrepancy can be due to two different factors, as we will see also in the following: a real mismatch in this region of the CMD between the observed and the simulated distributions, and thus a poor fit of the data using a SSP with a range of rotation velocities, or a wrong 
assumption for the distribution of rotation rates of the simulated objects.

\begin{figure}
\includegraphics[width=1\columnwidth]{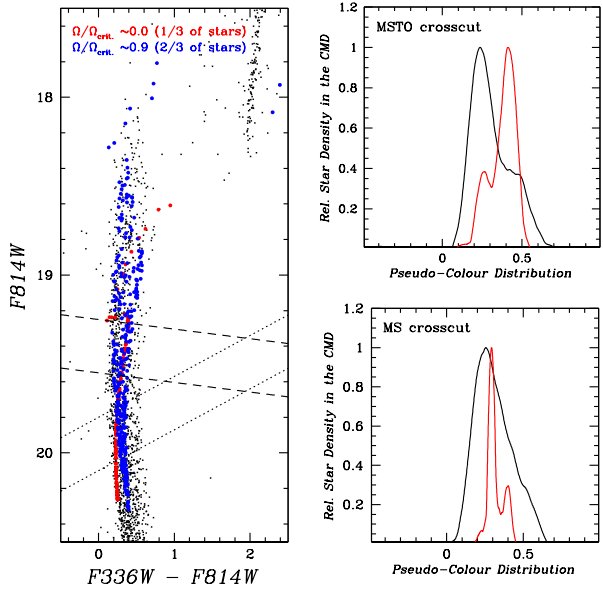}
\caption{Similar to Figure\,\ref{f:pseudo_rot_h10} but with overplotted on the cluster CMD a Monte Carlo simulation obtained using a single age population with a bimodal distribution for the rotation velocity rates. 2/3 of the stars are rapid rotators, $\Omega / \Omega_{\rm crit} = 0.9$, whereas 1/3 of the stars are slowly/non-rotating stars, $\Omega / \Omega_{\rm crit} \sim 0.0$. To provide a better understanding of what region of the CMD is occupied by the stars with different rotation rates we use a different colour for the two populations (non-rotating stars, red dots, fast-rotating stars, blue dots, respectively).}
\label{f:pseudo_rot_bimodal}
\end{figure}

In this context, we note that previous works in the Literature, aimed at reproducing the observed CMD morphology of young LMC clusters, in the context of the rotation velocities scenario, have adopted a bimodal distribution in the rotation rates instead of a continuous distribution as the one assumed here \citep[i.e.,][]{huang+10}. For example, \citet{dant+15} showed that the morphology of the MSTO in the young LMC star cluster NGC\,1856 can be interpreted as the combination of two stellar populations having the same age, composed for 2/3 of very rapid rotating stars ($\Omega / \Omega_{\rm crit} = 0.9$), and for 1/3 of slowly/non-rotating stars ($\Omega / \Omega_{\rm crit} \sim 0.0$). In this context, we test the possibility that the latter distribution also applies to our data. Hence, we derive a set of Monte Carlo simulations of a SSP, obtained using the same bimodal distribution and the same ratio as in \citet{dant+15}. The comparison between the observed and simulated CMD, and the pseudo-colour distributions are shown in Figure\,\ref{f:pseudo_rot_bimodal}. The pseudo-colour distributions clearly show that a bimodal distribution do not provide a good fit to this cluster data. In particular, the exclusion of stars at intermediate rotation velocities ($0.1 \leq \Omega / \Omega_{\rm crit} \leq 0.8$) skews the distribution towards fast-rotator stars and hence towards the red side of the pseudo-colour distribution in the MSTO crosscut, leaving the blue side heavily underpopulated.

\begin{figure}
\includegraphics[width=1\columnwidth]{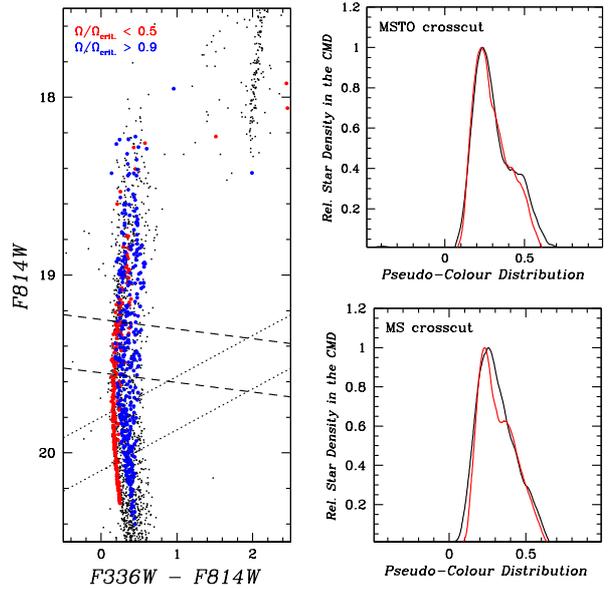}
\caption{Similar to Figure\,\ref{f:pseudo_rot_h10} but with overplotted on the cluster CMD a Monte Carlo simulation obtained by selecting from the simulation presented in Figure\,\ref{f:pseudo_rot_h10} only the stars with rotation velocities $\Omega / \Omega_{\rm crit} < 0.5$ and $\Omega / \Omega_{\rm crit}  > 0.9$. As in Figure\,\ref{f:pseudo_rot_bimodal} slow-rotators ($\Omega / \Omega_{\rm crit} < 0.5$) are reported in red, whereas fast-rotators ($\Omega / \Omega_{\rm crit} > 0.9$) are reported in blue.}
\label{f:pseudo_rot_best}
\end{figure}

As mentioned previously, another piece of evidence in favor of a bimodal distribution in the rotation velocity rates is the presence of a split in the MS of young LMC clusters \citep[see C17;][]{mari+18b}. Interestingly, Figure\,6 in C17 shows that the region in between the two splitted MSs coincides with the locus in the CMD populated by the isochrone with ``intermediate'' rotation velocities (i.e., $0.5 < \Omega / \Omega_{\rm crit} <  0.8$). In this context, we created a third simulation, selecting from the SYCLIST Monte Carlo simulation shown in Figure\,\ref{f:pseudo_rot_h10} only the stars with $\Omega / \Omega_{\rm crit} < 0.5$ and $\Omega / \Omega_{\rm crit} > 0.9$. Since the stars are extracted from the SYCLIST simulation described above, they still have an underlying ratio distribution that follows the one from \citet{huang+10}, that is each rotation velocity ``bin'' is not populated by the same number of stars and hence the distribution is not flat. Overall, the fraction of slow rotators with respect to the total number of stars is of the order of $\sim$ 40\%, slightly higher than the number derived in C17 from the analysis of the star distribution in the two splitted MS (i.e., 20-30\%).
This difference in the ratio between slow and fast rotators observed in NGC\,1831 and the younger clusters could be explained if, instead of a primordial mix of rotation ranges, we assume that fast-rotating stars evolve into slow/non-rotating stars, as suggested by \citet{dant+17}. The loss of angular momentum can be explained by a rapid breaking of the stellar layers that shifts the evolution of the stars from the fast-rotating track to the non-rotating track for the same stellar mass. This braking occurs in the core of the star and is due to the tidal interaction occurring in binary systems \citep[][and references therein]{kopa68,zahn77,dant+17}. A consequence of the braking process is that the time evolution of each ``braked'' star depends on the time at which braking is effective, that is the braked stars stop evolving along the fast-rotating isochrone and start evolving along the non-rotating one at different times. Hence, the pile-up of slowly-rotating stars braked at different ages can produce different populated MS and MSTO, depending among the other factors, on the cluster age. 

Figure\,\ref{f:pseudo_rot_best} shows the comparison between the observed CMD and the pseudo-colour distributions and the simulated ones. The resemblance between the pseudo-colour distributions is striking. The red side of the simulated MS is slightly underpopulated with respect to the observed one but overall both fits are very good.


Taking these results at face value, they suggest that the morphology of NGC\,1831 can be fully explained in the context of the rotation velocity scenario, under the assumption of a bimodal distribution of rotating stars with a negligible contribution from intermediate-velocity rotators (i.e., $0.5 < \Omega / \Omega_{\rm crit} < 0.9$). 

Finally, we note that our results are overall in good agreement with the ones obtained by G19. In fact, if we take into account only the eMSTO region, both scenarios (age spread and stellar rotation) provide a good fit to the data, preventing to provide conclusive evidence to exclude one of the two. In the caption of their Figure\, 6, G19 suggest that a model with only rotating stars  tends to favor a bimodal distribution of fast and slow rotators, with few intermediate rates, similarly to what we derive from our analysis. The slightly different ratio of rates between the two different analysis can be due to the different models adopted (SYCLIST vs MESA) and by the fact that in G19 the analysis was performed only in the eMSTO region, whereas in our case the choice of the distribution rates is such that both eMSTO and upper MS regions are fitted simultaneously. 


\section{Dynamical Analysis}
\label{s:dyn}

\begin{table*}
\begin{center}
\begin{tabular}{cccccccccc}
\hline
\hline
\multicolumn{1}{c}{Cluster} & \multicolumn{1}{c}{$V$} & \multicolumn{1}{c}{Aper.} & \multicolumn{1}{c}{Aper. Corr.} & \multicolumn{1}{c}{[Z/H]} & \multicolumn{1}{c}{$A_V$} & \multicolumn{1}{c}{$r_c$} & \multicolumn{1}{c}{$r_e$} & \multicolumn{2}{c}{log (${\cal{M}}_{\rm cl}/M_{\odot}$)}\\
(1) & (2) & (3) & (4) & (5) & (6) & (7) & (8) & (9) & (10)\\
 \hline
NGC\,1831 & 10.73 $\pm$ 0.07 & 50 & 0.22 $\pm$ 0.02 & -0.4 & 0.11 & 4.45 $\pm$ 0.15 & 8.55 $\pm$ 0.31 & 4.79 $\pm$ 0.10 & 4.55 $\pm$ 0.10\\
\hline 
\end{tabular}
\caption{Physical properties of the star cluster. Columns (1): Name of the cluster. (2): Integrated $V$ magnitude from \citet{goud+06}. (3): Radius in arcsec for the measure of the integrated magnitude. (4): Aperture correction in magnitude. (5): Metallicity (dex). (6): Visual extinction $A_V$ in magnitude. (7): Core radius $r_c$ in pc. (8): Effective radius $r_e$ in pc. (9-10): Logarithm of the cluster mass adopting a \citet{salp55} and a \citet{chab03} initial mass function,
respectively.}
\label{t:param}
\end{center}
\end{table*}

Although the analysis presented in Section\,\ref{s:simulation} seems to suggest that a rotation velocity scenario provides a better explanation for the observed feature in the CMD, in particular in the upper MS region, it is worth to verify if the presence of an age spread could still be plausible, maybe in combination with rates of different rotations, a possibility not excluded in the analysis presented by G19. In this context, one of the main conundrums in understanding the origin of the eMSTO in the MC intermediate-age star clusters is that they are not an ubiquitous feature, meaning that they are not present in all the clusters. \citet{goud+11a} suggested that the eMSTO can be hosted only by clusters where the escape velocities were higher than the wind velocities of the stars that are thought to provide the necessary material for secondary star formation, at the time such stars, defined as first-generation polluters, were present in the cluster (the so-called ``early escape velocity scenario''). The main candidate stars as ``polluters'' are: {\it (i)\/} AGB stars with
$4 \la {\cal{M}}/M_{\odot} \la 8$ \citep{danven08}, {\it (ii)\/} fast rotating massive stars \citep[``FRMS'',][]{decr+07}. and {\it (iii)\/} massive binary stars \citep{demink+09}. Subsequently, \citet{goud+14} and \citet{corr+14}, from their studies of a sample of clusters with and without eMSTOs, suggested that the critical escape velocity threshold above which the cluster is able to retain the material ejected by the first-generation polluters listed above is in the range of 12\,--\,15 \kms. Hence, here we want to verify whether NGC\,1831 possesses the right physical and dynamical properties that allow the retention of the polluters ejecta and thus the formation of second-generation stars. In this context, in the following sections we determine the structural and
dynamical parameters of the clusters, following the methodology described in detail in our previous works.   

\subsection{Structural Parameters}
\label{s:king}

\begin{figure}
\centerline{\includegraphics[width=0.8\columnwidth]{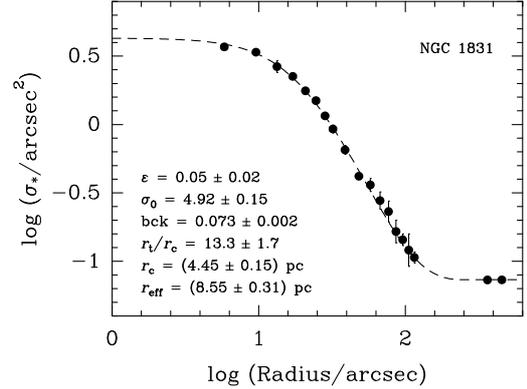}}
\caption{Radial surface number density profile of NGC\,1831. Black points represent the observed values, whereas the dashed line represents the best-fit King model (Eq.~\ref{eq:King}) whose parameters are shown in the legend.}
\label{f:king}
\end{figure}

After deriving completeness-corrected surface number densities of stars in the images following the procedure described in detail in \citet{goud+09}, we fit the radial surface number density profile using a \citet{king62} model combined with a constant background level, adopting the following equation:
\begin{equation}
\sigma(r) = \sigma_0 \: \left( \frac{1}{\sqrt{1 + (r/r_c)^2}} - \frac{1}{\sqrt{1+c^2}}
\right)^2 \; + \; {\rm bkg} 
\label{eq:King}
\end{equation}
where $\sigma_0$ is the central surface number density, $r_c$ is the core radius,  $c
\equiv r_t/r_c$ is the King concentration index ($r_t$ being the tidal radius),
and $r$ is the equivalent radius of the ellipse ($r = a\,\sqrt{1-\epsilon}$,
where $a$ is the semi-major axis of the ellipse and $\epsilon$ is its
ellipticity). The background level is derived exploiting the ACS parallel field located $\sim$ 6\arcmin\ away from the cluster center, using the same limiting {\em F814W} magnitude as in the WFC3 image. Figure\,\ref{f:king} shows the best-fit model, derived using a $\chi^2$ minimization routine. The derived core radius is $r_c$ = 18\farcs3 $\pm$ 0\farcs6, which corresponds to 4.45 $\pm$ 0.15 pc, whereas the effective radius $r_e$ is 35\farcs2 $\pm$ 1\farcs4 (8.55 $\pm$ 0.31 pc), which is consistent with the literature estimate to within 1$\sigma$ \citep[i.e., $r_e$ = 8.22 pc,][]{mclvan05}. 

\subsection{Cluster mass and escape velocity}
\label{s:escvel}

The mass and escape velocity of the cluster are calculated as a function of time, going back to an age of 10 Myr. This age limit represents the time at which the cluster has survived the era of violent relaxation and when the most massive stars of the first generation (i.e., FRMS and massive binaries), considered in the literature as potential candidate polluters, are expected to start losing mass through slow winds. 

We determine the current cluster mass from its integrated {\it V}-band magnitude, taken from \citet{goud+06} and listed in Table\,\ref{t:param}. As a first step we derive the aperture correction for the {\it V} magnitude from the best-fit King model, calculating the fraction of the total cluster light that is encompassed by the measurement aperture (in this case a 50$''$ radius). Then, after applying the aperture correction, we derive the cluster mass from the values of $A_V$, $(m-M)_0$, and $[Z/H]$ reported in Table\,\ref{t:param}, interpolating between the $M/L_V$ values in the SSP model tables of \citet{brucha03}, assuming a \citet{salp55} initial mass function. 

The escape velocity of the cluster and its evolution with time is derived using the prescriptions of \citet{goud+14}. The evolution of the cluster mass and radius are evaluated assuming both the absence/presence of mass segregation, which plays a crucial role in terms of the evolution in the early stage of the cluster's expansion and mass-loss rate \citep[e.g.,][]{mack+08b,vesp+09}. As in C15 and C17, for the cluster model with initial mass segregation, we adopt the results from the simulation named SG-R1 in \citet{derc+08}, which involves a
model cluster that features a level of initial mass segregation of
$r_e/r_{e,>1}$ = 1.5, where $r_{e,>1}$ is the effective radius of the cluster for stars with ${\cal{M}} >$ 1 \Msun\, \citep[see][for a detailed description about this choice]{goud+14}. Finally, escape velocities are obtained from the reduced gravitational potential $V_{\rm esc} (r,t) = (2\Phi_{\rm tid} (t) - 2\Phi (r,t))^{1/2}$, at the core radius. The choice to derive the escape velocity at the cluster core radius is in conformity with the prediction of the {\it in situ\/} scenario \citep{derc+08}, in which the second generation stars are formed in the innermost region of the cluster. The current cluster escape velocity is 8.2 $\pm$ 0.4 \kms\ at the effective radius and 11.03 $\pm$ 0.5 \kms\, at the core radius, assuming a \citet{salp55} initial mass function (IMF), whereas the values change to 6.2 $\pm$ 0.3 \kms\, and 8.3 $\pm$ 0.4 \kms, respectively, for the \citet{chab03} IMF.

\begin{figure}
\centerline{\includegraphics[width=0.6\columnwidth]{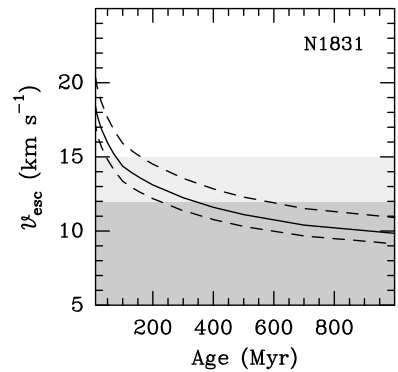}}
\caption{The solid line represents the derived escape velocity $V_{\rm esc}$ as function of time for NGC 1831. Dashed lines indicate $\pm 1 \sigma$ uncertainties. The light gray region represents the range for $V_{\rm esc}$ of 12\,--\,15 \kms\ described in Section~\ref{s:escvel}. The dark gray region indicates $V_{\rm esc} \leq 12$ \kms,  believed to be too small to retain material shed by massive stars and AGB stars.}
\label{f:escvel}
\end{figure}

In Figure\,\ref{f:escvel}, we show the escape velocity as a function of time. In particular, we show the ``plausible'' escape velocity derived using the approach described in detail in \citet{goud+14} and adopted in C17, which involves a procedure that takes into account the various results from the compilation of MC star cluster properties and N-body simulations by \citet{mack+08b}. In Figure\,\ref{f:escvel} we also highlight the critical escape velocity range of 12\,--\,15 \kms\, (depicted as a light gray region), and the region below 12 \kms\, that represents the range in which eMSTOs were not observed in LMC star clusters by \citet{milo+09} and \citet{corr+14}. 

Interestingly, we find $V_{\rm esc} = 18.4 \pm 1.9$ \kms\ for NGC 1831 at an age of 10 Myr, and $V_{\rm esc}$ stays above 15 \kms\ for about 100 Myr. The threshold of 15 \kms\ was suggested by \citet{goud+14} to be the approximate limit below which clusters cannot retain sufficient amounts of material shed by winds of first-generation stars to allow the creation of second-generation stars. Since we are able to explain the morphology of the wide upper MS and MSTO in NGC 1831 by the effects of stellar rotation without having to invoke an additional range of stellar ages, the current results suggest that any threshold for $V_{\rm esc}$ to allow retention of stellar mass-loss material is larger than 15 \kms. Looking at the results for clusters whose eMSTO morphology could not easily be explained with only stellar rotation \citep[e.g.,][]{goud+14,goud+17,milo+17,cost+19}, we suggest that this threshold would have to be at least 20 \kms. 

\section{Summary and Conclusion}
\label{s:summary}
 
We present the results of a study, performed analysing deep WFC3 {\it HST} images, of the wide MS and MSTO observed in the $\sim$ 800 Myr old star cluster NGC\,1831.

We compare the cluster CMD morphology with Monte Carlo simulations of synthetic star clusters with multiple SSP of different ages and with a single SSP with a range of different rotation velocities in order to address whether a given scenario (age spread vs stellar rotation) can better reproduce the observed CMD. We find that in the first scenario, the simulation provides a very good fit of the eMSTO, but fails to reproduce the full colour width of the MS. For what concerns the second scenario, we find that a range of stellar rotation with the same distribution as in \citet{huang+10} or assuming a bimodal distribution as in \citet{dant+15}, that is with 1/3 of non-rotating/slowly rotating stars ($\Omega / \Omega_{\rm crit} \sim 0.0$) and 2/3 of fast rotating stars ($\Omega / \Omega_{\rm crit} \sim 0.95$) cannot reproduce the data. However, we derive that if we assume a bimodal distribution that excludes intermediate rotation velocities (0.5 < $\Omega / \Omega_{\rm crit} < 0.9$), we obtain a simultaneous good fit of both the MS and MSTO. With this assumption, we obtain that the ratio of the slow rotating stars with respect to the total is $\sim$ 40\%. Similar bimodal distributions have been observed also in younger clusters (e.g., NGC\,1850, C17), although with a sligthly lower ratio of slow rotators ($\sim$ 20\,--\,30\%). These results seem to suggest that the origin of slow-rotating and non-rotating stars can be attributed to the tidal interaction in binary systems, as suggested by \citet{dant+15,dant+17}, more than to the loss of angular momentum during the pre-MS phase due to magnetic breaking \citep{zorroy12}. In fact, in the first hypothesis the time scale of the braking depends on both the stellar mass and its evolutionary stage, and on the parameters of the binary system, so braked stars may be present in clusters over a wide range of ages. Moreover, the braking process can be effective at different times for the stars involved in the process,  and the ratio of slow-to-fast rotators can vary with the time evolution and hence with the age of the clusters.

Finally, we derive the dynamical properties of NGC\,1831 and we obtain that the cluster had an escape velocity $V_{\rm esc} = 18.4 \pm 1.9$ \kms\ at an age of $\sim$ 10 Myr and that $V_{\rm esc}$ stays above 15 \kms\, for about 100 Myr. Hence, in principle $V_{\rm esc}$ could be high enough to retain the material shed by the slow winds of the first-generation ``polluter stars''.   

These results confirm the fundamental role played by stellar rotation in shaping the CMD morphology of some LMC young and intermediate-age star clusters, without the need to invoke the presence of an age spread. They also indicate that, for the clusters in which stellar rotation alone cannot fully explain the nature of the eMSTO phenomenon, the escape velocity threshold should have to be higher (at least $\sim$ 20 \kms) than what was previously suggested \citep[][and references therein]{corr+14,goud+14}.

\section*{Acknowledgments}
Support for this project was provided by NASA through grant HST-GO-14688 from the Space Telescope Science Institute, which is operated by the Association of  Universities for Research in Astronomy, Inc., under NASA contract NAS5--26555. We made significant use of the SAO/NASA Astrophysics Data System during this project. 

\section*{Data availability}
MAST data underlying this article are available at doi: http://dx.doi.org/10.17909/t9-22ff-gs44
Final data products (i.e., catalogs) underlying this article will be shared on reasonable request to the corresponding author.


\label{lastpage}


\begin{thebibliography}{99}


\bibitem[Bastian \& de Mink(2009)]{basdem09}
Bastian, N., \& de Mink, S.~E. 2009, MNRAS, 398, L11

\bibitem[Bastian et al.(2017)]{bast+17}
Bastian, N., Cabrera-Ziri, I, Niederhofer, F., et al. 2017, MNRAS, 465, 4795

\bibitem[Bastian et al.(2018)]{bast+18}
Bastian, N., Kamann, S., Cabrera-Ziri, I., et al. 2018, MNRAS, 480, 3793

\bibitem[Bellini \& Bedin(2009)]{belbed09}
Bellini, A., \& Bedin, L. R. 2009, PASP, 121, 1419

\bibitem[Bellini et al.(2011)Bellini, Anderson, \& Bedin]{bell+11}
Bellini, A., Anderson, J., \& Bedin, L. R. 2011, PASP, 123, 622

\bibitem[Bellini et al.(2014)]{bell+14}
Bellini, A., Anderson, J., van der Marel, R. P., et al. 2014, ApJ, 797, 115

\bibitem[Bellini et al.(2017)]{bell+17}
Bellini, A., Anderson, J., Bedin, L. R., et al. 2017, ApJ, 842, 6

\bibitem[Brandt \& Huang(2015)]{brahua15}
Brandt, T. D., \& Huang, C. X. 2015, ApJ, 807, 24

\bibitem[Bressan et al.(2012)]{bres+12}
Bressan, A., Marigo, P., Girardi, L., et al. 2012, MNRAS, 427, 127

\bibitem[Bruzual \& Charlot(2003)]{brucha03}
Bruzual, G. A., \& Charlot, S. 2003, MNRAS, 379, 1431

\bibitem[Chabrier(2003)]{chab03}
Chabrier, G.\ 2003, PASP, 115, 763

\bibitem[Claret(2000)]{clar00}
Claret, A. 2000, A\&A, 363, 1081

\bibitem[Cordoni et al.(2018)]{cord+18}
Cordoni, G., Milone, A. P., Marino, A. F., et al. 2018, ApJ, 869, 139

\bibitem[Correnti et al.(2014)]{corr+14}
Correnti, M., Goudfrooij, P., Kalirai, J. S., et al. 2014, ApJ, 793, 121

\bibitem[Correnti et al.(2015)]{corr+15}
Correnti, M., Goudfrooij, P., Puzia, T. H., \& de Mink, S. E. 2015, MNRAS, 450, 3054 (C15)

\bibitem[Correnti et al.(2017)]{corr+17}
Correnti, M., Goudfrooij, P., Bellini, A., Kalirai, J. S., \& Puzia, T. H. 2017, MNRAS, 467, 3628 (C17)

\bibitem[Costa et al.(2019)]{cost+19}
Costa, G., Girardi, L., Bressan, A., et al.\ 2019, \aap, 631, A128

\bibitem[D'Antona \& Ventura(2008)]{danven08}
D'Antona, F., \& Ventura, P. 2008, MNRAS, 479, 805

\bibitem[D'Antona et al.(2015)]{dant+15}
D'Antona, F., Di Criscienzo, M., Decressin, T., et al. 2015, MNRAS, 543, 2637 

\bibitem[D'Antona et al.(2017)]{dant+17}
D'Antona, F., Milone, A. P., Tailo, M., et al. 2017, Nature Astronomy, 1, 0186 

\bibitem[D'Antona et al.(2018)]{dant+18}
D'Antona, F., Milone, A. P., Tailo, M. et al. 2018, Mmsait, 89, 42

\bibitem[Decressin et al.(2007)]{decr+07}
Decressin, T., Meynet, G., Charbonnel, C., Prantzos, N., \& Ekstr\'om, S.
2007, A\&A, 464, 1029

\bibitem[de Mink et al.(2009)]{demink+09}
de Mink, S. E., Pols, O. R., Langer, N., \& Izzard, R. G. 2009, A\&A, 5007, L1

\bibitem[D'Ercole et al.(2008)]{derc+08}
D'Ercole, A., Vesperini, E., D'Antona, F., McMillan, S. L. W., \& Recchi, S. 2008, MNRAS, 391, 825

\bibitem[Dupree et al.(2018)]{dupr+18}
Dupree, A. K., Dotter, A., Johnson, C. I., et al. 2018, ApJL, 846, L1

\bibitem[Ekstr\"{o}m et al.(2012)]{ekst+12}
Ekstr\"{o}m, S., Georgy, C., Eggenberger, P., et al. 2012, A\&A, 537, 146

\bibitem[Espinosa Lara \& Rieutord(2011)]{esplar11}
Espinosa Lara, F., \& Rieutord, M. 2011, A\&A, 533, 43

\bibitem[Georgy et al.(2013)]{geor+13}
Georgy, C., Ekstr\"{o}m, S., Granada, A., et al. 2013, A\&A, 553, 24

\bibitem[Georgy et al.(2014)]{geor+14}
Georgy, C., Granada, A., Ekstr\"{o}m, S., et al. 2014, A\&A, 566, 21

\bibitem[Georgy et al.(2019)]{geor+19}
Georgy, C., Charbonnel, C., Amard, L., et al. 2019, A\&A, 622, A66

\bibitem[Gilliland(2004)]{gill04}
Gilliland, R. L. 2004, ACS/ISR 2004-01 (Baltimore, MD: STScI), available online at http://www.stsci.edu/hst/acs/documents/isrs

\bibitem[Gilliland et al.(2010)Gilliland, Rajan, \& Deustua]{gill+10}
Gilliland, R. L., Rajan. A., \& Deustua, S. 2010, WFC3/ISR 2010-10 (Baltimore, MD: STScI), available online at http://www.stsci.edu/hst/efc3/documents/isrs

\bibitem[Girardi et al.(2009)Girardi, Rubele, \& Kerber]{gira+09}
Girardi, L., Rubele, S., \& Kerber, L. 2009, MNRAS, 394, L74

\bibitem[Girardi et al.(2011)Girardi, Eggenberger, \& Miglio]{gira+11}
Girardi, L., Eggenberger, P., \& Miglio, A. 2011, MNRAS, 412, L10

\bibitem[Girardi et al.(2013)]{gira+13}
Girardi, L., Goudfrooij, P., Kalirai, J. S., et al. 2013, MNRAS, 431, 3501

\bibitem[Gonzaga et al.(2012)]{gonz+12}
Gonzaga, S., Hack, W., Fruchter, A. S., \& Mack, J., eds. 2012, The DrizzlePac Handbook (Baltimore, STScI) 

\bibitem[Gossage et al.(2019)]{goss+19}
Gossage, S., Conroy, C., Dotter, A., et al. 2019, ApJ, 887, 199 (G19)

\bibitem[Goudfrooij et al.(2006)]{goud+06}
Goudfrooij, P., Gilmore, D., Kissler-Patig, M.,  Maraston C.\ 2006, \mnras, 369, 697

\bibitem[Goudfrooij et al.(2017)Goudfrooij, Girardi, \& Correnti]{goud+17}
Goudfrooij, P., Girardi, L., \& Correnti, M. 2017, 846, 22

\bibitem[Goudfrooij et al.(2009)]{goud+09}
Goudfrooij, P., Puzia, T. H., Kozhurina-Platais, V., \& Chandar, R. 2009,
AJ, 137, 4988

\bibitem[Goudfrooij et al.(2011a)]{goud+11a}
Goudfrooij, P., Puzia, T. H., Chandar, R., \& Kozhurina-Platais, V. 2011, ApJ, 737, 4
\bibitem[Goudfrooij et al.(2014)]{goud+14}
Goudfrooij, P., Girardi, L., Kozhurina-Platais, V., et al. 2014, ApJ, 797, 35

\bibitem[Goudfrooij et al.(2015)]{goud+15}
Goudfrooij, P., Girardi, L., Rosenfield, P., et al. 2015, MNRAS, 450, 1693

\bibitem[Goudfrooij et al.(2018)]{goud+18}
Goudfrooij, P., Girardi, L., Bellini, A., et al. 2018, ApJ, 864, L3 (G18)

\bibitem[Huang et al.(2010)Huang, Gies, \& McSwain]{huang+10}
Huang, W., Gies, D. R., \& McSwain, M. V. 2010, Apj, 722, 605

\bibitem[Kamann et al.(2020)]{kama+20}
Kamann, S., Bastian, N., Gossage, S. et al. 2020, MNRAS, 492, 2177

\bibitem[Kopal(1968)]{kopa68}
Kopal, Z. 1968, Ap\&SS, 1, 179 

\bibitem[King(1962)]{king62}
King, I. 1962, AJ, 67, 471

\bibitem[Mackey et al.(2008a)]{mack+08a}
Mackey, A. D., Broby Nielsen. P., Ferguson, A. M. N., \& Richardson,
 J. C. 2008, ApJ, 681, L17

\bibitem[Mackey et al.(2008b)]{mack+08b}
Mackey, A. D., Wilkinson, M. I., Davies, M. B., \& Gilmore, G. F. 2008, MNRAS, 386, 65

\bibitem[Marino et al.(2018a)]{mari+18a}
Marino, A. F., Milone, A. P., Casagrande, L., et al. 2018a, ApJ, 863, L33

\bibitem[Marino et al.(2018b)]{mari+18b}
Marino, A. F., Przybilla, N., Milone, A. P., et al. 2018b, AJ, 156, 116

\bibitem[McLaughlin \& van der Marel(2005)]{mclvan05}
McLaughlin, D. E., \& van der Marel, R. P. 2005, ApJs, 161, 304

\bibitem[Milone et al.(2009)]{milo+09}
Milone, A. P., Bedin, L. R., Piotto, G., \& Anderson, J.\ 2009, A\&A, 497, 755

\bibitem[Milone et al.(2015)]{milo+15}
Milone, A. P., Bedin, L. R., Piotto, G., et al. 2015, MNRAS, 450, 3750

\bibitem[Milone et al.(2016)]{milo+16}
Milone, A. P., Marino, A. F., D'Antona, F., Bedin, L. R., Da Costa, G. S., Jerjen, H., \& Mackey, A. D. 2016, MNRAS, 458, 4368

\bibitem[Milone et al.(2017)]{milo+17}
Milone, A. P., Marino, A. F., D'Antona, F., et al. 2017, MNRAS, 465, 4363

\bibitem[Milone et al.(2018)]{milo+18}
Milone A. P., Marino, A. F., Di Criscienzo, M., et al. 2018, MNRAS, 477, 2640

\bibitem[Mowlavi et al.(2012)]{mowl+12}
Mowlavi, N., Eggenberger, P., Meynet, G., et al. 2012, A\&A, 541, 41

\bibitem[Niederhofer et al.(2015)]{nied+15}
Niederhofer, F., Hilker, M., Bastian, N., \& Silva-Villa, E. 2015, A\&A, 575, 62

\bibitem[Rubele et al.(2011)]{rube+11}
Rubele, S., Girardi, L., Kozhurina-Platais, V., Goudfrooij, P., \& Kerber, L.
 2011, MNRAS, 414, 2204

\bibitem[Salpeter(1955)]{salp55}
Salpeter, E. E. 1955, ApJ, 121, 161

\bibitem[Silverman(1986)]{silv86} 
Silverman, B. W. 1986, in {\it Density Estimation for Statistics and Data Analysis}, Chap and Hall/CRC Press, Inc. 

\bibitem[Vesperini et al.(2009)Vesperini, McMillan, \& Portegies Zwart]{vesp+09}
Vesperini, E., McMillan, S. L. W., \& Portegies Zwart, S.\ 2009, ApJ, 698, 615

\bibitem[Zahn(1977)]{zahn77}
Zahn, J. P. 1977, A\&A, 57, 383

\bibitem[Zorec \& Royer(2012)]{zorroy12}
Zorec, J., \& Royer, F. 2012, A\&A, 537, 120 


\end{thebibliography}
\end{document}